\documentclass[preprint,aps,floatfix,superscriptaddress,prb]{revtex4}

\usepackage{amsmath, amsthm, amssymb, graphicx}
\usepackage{color}
\usepackage{multirow}
\usepackage{ulem}
\usepackage{float}
\usepackage{bm}
\usepackage{subfiles}
\usepackage{siunitx}
\usepackage{booktabs}
\usepackage{array}

\begin{document}

\title{Late stages in the ordering of magnetic skyrmion lattices}
\author{James Stidham}
\affiliation{Department of Physics, Virginia Tech, Blacksburg, VA 24061-0435, USA}
\affiliation{Center for Soft Matter and Biological Physics, Virginia Tech, Blacksburg, VA 24061-0435, USA}
\author{Michel Pleimling}
\affiliation{Department of Physics, Virginia Tech, Blacksburg, VA 24061-0435, USA}
\affiliation{Center for Soft Matter and Biological Physics, Virginia Tech, Blacksburg, VA 24061-0435, USA}
\affiliation{Academy of Integrated Science, Virginia Tech, Blacksburg, VA 24061-0563, USA}
\date{\today}

\begin{abstract}
The late-stage ordering of interacting magnetic skyrmions is studied numerically through extensive Langevin molecular dynamics simulations.
Defining skyrmion displacements that change the connectivity of cells obtained in a Voronoi tesselation as events, we investigate 
event histograms as a function of the time elapsed since preparing the system as well as the histograms of consecutive events as a function of the
time separating these two events. These histograms, which provide unique insights into the transient properties 
during the ordering process of skyrmion matter, show a characteristic behavior that allows the
Magnus-force dominated regime, where the Magnus force accelerates the relaxation process, to be distinguished from the noise-dominated regime,
where the Magnus force enhances the effects of thermal noise. In the Magnus-force dominated regime the different histograms display
power-law tails with exponents that depend on the strength of the Magnus force.
\end{abstract}

\maketitle

\section{Introduction}
Since their first experimental realizations \cite{Muh09,Yu10} and, especially, since their observation at room temperature \cite{Woo17,Bou16,Des19}, 
arrangements of magnetic skyrmions, nanometer sized particlelike spin textures \cite{Ros06,Nag13}, have attracted much attention \cite{Bac20}.
Racetrack memory devices \cite{Fer13,Tom15,Vak20} and logic gates \cite{Zha15} are some of the possible promising applications of
these spin textures.

One of the characteristic features of skyrmions is the presence of the nondissipative Magnus force. As this force acts perpendicular
to the skyrmion's velocity, it has a major effect on the dynamics of the spin textures, especially in cases where skyrmions are not isolated,
but interact with each other and/or with defects. Intriguing new phenomena can emerge under these circumstances. For example, in driven systems a constant angle,
known as the skyrmion Hall angle, appears between the direction of the drive and the direction of displacement \cite{Nag13,Rei15,Rei16,Dia17,Jia17,Lit17,Jug19,Viz20}. 
Other studies have focused on the appearance of avalanches when the driven skyrmions interact with random pins \cite{Dia18}, the pattern formation
of geometrically confined skyrmions \cite{Sch19}, the existence of nonequilibrium phase transitions between different dynamic phases 
\cite{Rei18,Bro19}, shear banding of driven skyrmions in inhomogeneous pinning arrays \cite{Rei20}, laning transitions 
in presence of damping \cite{Rei19}, as well as the formation of skyrmion crystals \cite{Kim20,Nak17} and the flow of skyrmion lattices \cite{Sat19}.

Intriguingly, randomly nucleated skyrmions may be subjected to thermal fluctuations \cite{Mil18,Zha19,Noz19} and undergo a relaxing process.
Some aspects of the ordering of interacting
skyrmions has been investigated through Langevin molecular dynamics \cite{Brown18,Brown19} simulations and has also been observed recently in experiments
using sub-nm thick CoFeB-based multilayer systems \cite{Zaz20}.
These studies have revealed that the interplay between repulsive skyrmion-skyrmion interaction, Magnus force, and thermal noise yields different dynamic
regimes. In the Magnus-force dominated regime that prevails for weak thermal noise, the Magnus force cooperates with the repulsive skyrmion-skyrmion
interaction and accelerates the relaxation to the triangular lattice. Strong thermal noise yields the noise-dominated regime that is characterized by
the fact that the Magnus force amplifies the effects of the thermal noise so that the dynamics is noisier than in absence of this
nondissipative force. Even richer scenarios are possible in situations where the mobility of skyrmions is affected by pinning that stems from material
imperfections \cite{Liu13,Iwa13,Sam13,Mul15,Kim17,Leg17}. Whereas collective pinning due to strong attractive defects can yield a skyrmion glass \cite{Hos18},
pinning is reduced for weaker pinning strengths as the Magnus force helps skyrmions avoid caging effects by bending around defects \cite{Brown19}.

The aim of the present work is to provide an in-depth view of the ordering process of interacting skyrmions in situations dominated
by the Magnus force and / or thermal noise. Whereas in Ref. \cite{Brown18} the emphasis was on the early stages of the 
ordering process, including a possible aging scaling regime,
in this study we also investigate the later stages where only a few defects persist in an otherwise well ordered skyrmion lattice. 
Our study, which to our knowledge encompasses the first systematic numerical investigation of the relaxation processes in the late-time 
dynamic regime before the system reaches equilibrium, highlights the usefulness of event statistics in the context of skyrmion matter.
In order to monitor
this ordering process, we consider at every timestep a Voronoi tesselation of the skyrmion system and count the number of cells that have six edges. We then define
as events skyrmion displacements that change the number of cells with six edges. In the late stages of the process, these events are rare and 
a large amount of timesteps separate consecutive events. Our analysis is based on histograms of the number of events that have taken place since preparing
the system as well as on histograms of the number of consecutive events that take place within a given time interval. We show that these histograms 
not only allow us to identify the different dynamic regimes, they also are characterized in the Magnus-force dominated regime by algebraic tails governed
by exponents whose values depend on the strength of the Magnus force. For weak noise these exponents are found to not depend on the noise strength.

In the next Section, we first review the particle-based model \cite{Lin13} used for our Langevin molecular dynamics simulations. We also discuss 
in more detail the quantities that we investigate in our work. In Section III, we first present results without thermal noise, thus focusing
exclusively on the Magnus-force dominated regime. We then add thermal noise and show that this leads to different dynamic regimes. The final Section
presents our conclusions.

\section{Model and quantities}
We follow previous work \cite{Brown18} and consider a particle-based model \cite{Lin13} of interacting skyrmions moving on
a two-dimensional surface. The rigid structure approximation underlying the model is valid in the low-density regime
\cite{Nag13,Jia17,Pollath17}.
In this description, skyrmions interact via a long-range repulsive force and are subjected to the velocity-dependent Magnus force.
The skyrmions can be driven by an electrical field and can interact with defects. In this paper, we focus on ordering processes
taking place in the absence of an external drive and in situations where skyrmion-defect interactions can be omitted.
Following Thiele's approach \cite{Thiele73}, the system is then modeled as a set of Langevin equations for the skyrmion drift
velocities:
\begin{equation} \label{eq:langevin}
\eta {\bf v}_i = \beta \hat{z} \times {\bf v}_i + {\bf F}_i^s + {\bf  f}
\end{equation}
where the $N_s$ different skyrmions are labeled by the index $i = 1, \cdots, N_s$, whereas $\eta$ is the damping coefficient.
The first term on the right hand side is the nondissipative Magnus force with strength $\beta$, acting in the direction
perpendicular to the velocity. The presence of this force, which results in curved trajectories, has a significant impact
on the dynamic properties of interacting skyrmions, especially far from stationarity \cite{Brown18,Brown19}.
We impose the constraint $\eta^2 + \beta^2 = 1$ from which follows that the average velocity magnitude of a 
free skyrmion is not dependent on the Magnus force. We focus on the following values of the ratio $\alpha = \beta / \eta$ of the Magnus force 
strength and the damping coefficient: $\alpha = 0$, i.e. no Magnus force, $\alpha = 5$, and $\alpha =
9.962$, which is a realistic value for MnSi \cite{Lin13,Reichhardt15}.
${\bf F}_i^s$ denotes the resultant force on skyrmion $i$ due to the presence of the other skyrmions: ${\bf F}_i^s =
\sum\limits_{j \ne i} K_1(r_{ij}/\xi) \hat{\bf r}_{ij}$ with the modified Bessel function of the second kind $K_1$ \cite{Lin13},
the healing length $\xi$ (we choose units such that the healing length $\xi = 1$),
and the unit vector $\hat{\bf r}_{ij} = {\bf r}_{ij}/r_{ij} = ( {\bf r}_i - {\bf r}_j ) / | {\bf r}_i - {\bf r}_j|$ pointing
from skyrmion $j$ to skyrmion $i$. For large distances, $r_{ij}/\xi \gg 1$, this force decays exponentially. In our
Langevin dynamics simulations we use a cut-off length $\lambda$ with typically $\lambda = 7 \xi$ (see also inset in Fig. \ref{fig3}). Finally, the last term in Eq. 
(\ref{eq:langevin}) describes delta-correlated thermal white noise with mean 0: $\langle f_\mu (t) \rangle = 0$ and
$\langle f_\mu (t) f_\nu (t') \rangle = \sigma^2 \delta_{\mu \nu} \delta(t - t')$, with $\sigma^2 = 2 \eta k_B T$.

The two-dimensional systems we discuss in the following typically have the size $\frac{2}{\sqrt{3}} L \times L$ with $L=72$
in units of skyrmion radii,
but we also considered systems with $L=36$ in order to discuss finite-size effects. The combination of this
shape and periodic boundary conditions allow the skyrmions in absence of thermal noise to settle into a triangular lattice at equilibrium.
The skyrmion density we choose was $10\%$, which results in systems containing 596 skyrmions.

\begin{figure}
 \centering \includegraphics[width=0.6\columnwidth,clip=true]{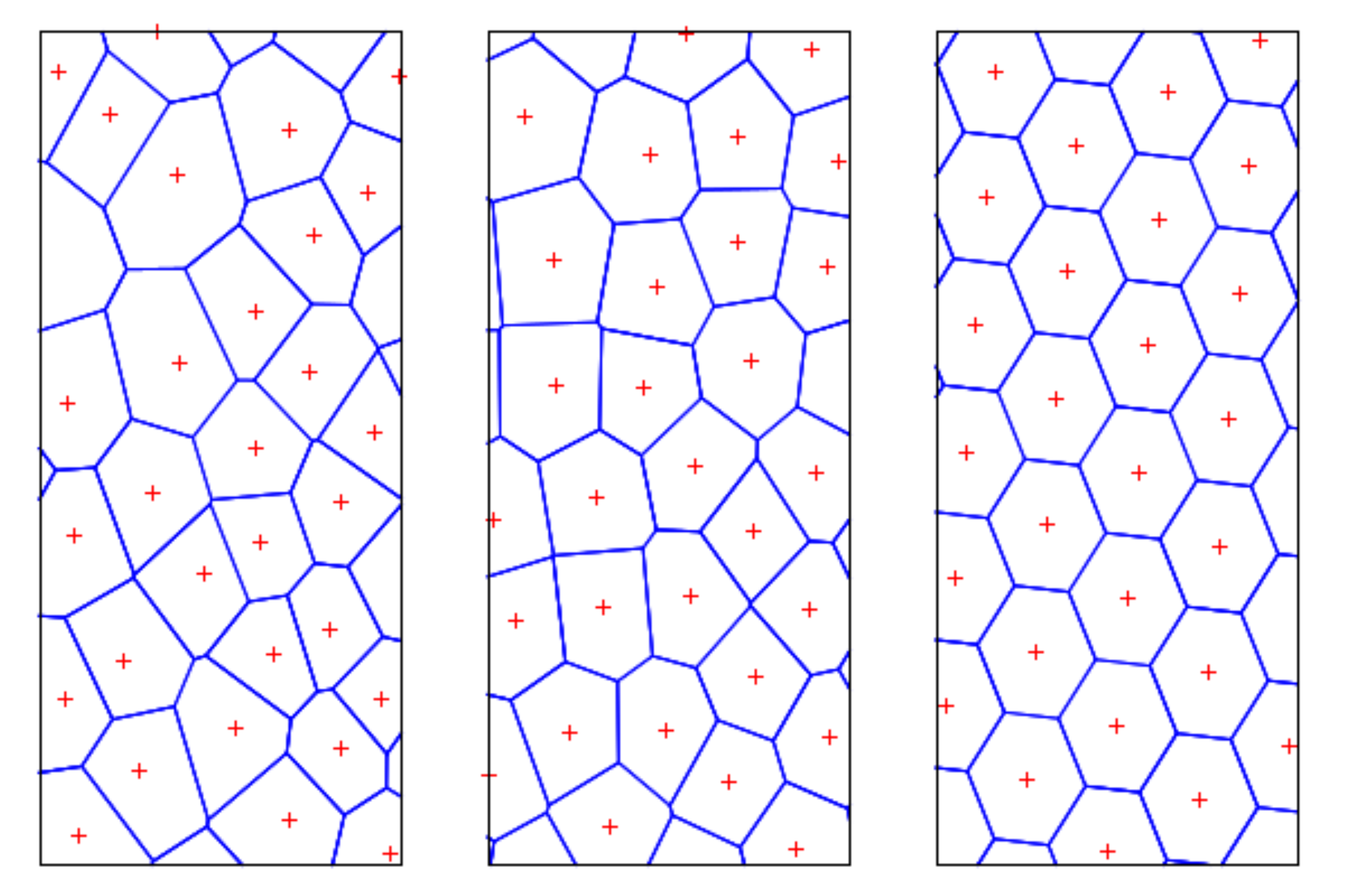}
\caption{Part of a skyrmion lattice at different times since preparing the system in a disordered initial state.
From left to right: $t=100$, $t=1,000$, and $t=10,000$. The skyrmion positions are indicated by the red marks. The lines
result from the Voronoi tesselations. In the perfect orderered triangular lattice all polygons have six edges.
The values of the system parameters are $L=36$, $\lambda = 7 \xi$, $\alpha = 9.962$, and $\sigma = 0$.
}
\label{fig1}
\end{figure}

Initially, at time $t=0$, we randomly distribute the skyrmions in the domain, where we make sure that skyrmions are not overlapping.
The equations of motion (\ref{eq:langevin}) are then solved with a standard fourth-order Runge-Kutta method and the integration
timestep $dt = 0.01$. In the absence of external drive and defects that can pin the skyrmions, the skyrmions undergo an ordering
process and relax towards an (in presence of weak noise approximately) ordered triangular lattice. 
In the absence of noise and in the large-volume limit the system will not display perfect long-range order, in agreement with the
Hohenberg-Mermin-Wagner theorem. However, our systems are so small that a hexatic phase can not be distinguished from a crystalline
phase. A convenient way to assess
the level of ordering is through Voronoi tesselation that readily reveals local ordering in an (approximate) hexagonal structure
through the appearance of six-sided cells, whereas in regions with defects in the hexagonal structure, cells with more or less
than six edges persist \cite{Brown18,Brown19}. Starting from a disordered state, the Voronoi tesselation initially yields a mixture of oddly-shaped
cells with variable numbers of edges. As time progresses and the skyrmions start to order in a hexagonal structure, six-sided cells
quickly prevail, with the exception of long lasting defects that manifest themselves through persisting pairs of cells
where one has five and the other seven edges.
The configurations shown in Fig. \ref{fig1} illustrate this relaxation process.

%####################### Figure 1 ############################%
%####################### Figure 1 ############################%

In relaxation and ordering processes different dynamic regimes can be identified \cite{Henkel10} when comparing the time
elapsed since preparing the system with both the microscopic time scale $t_{micro}$ and the equilibration time $t_{eq}$.
In the early regime, with $t \gtrsim t_{micro}$, the system remains in quasi-equilibrium in a state close to the state of initial
preparation before passing into the aging regime. In the aging or intermediate regime, times are much larger than microscopic
times, but much smaller than the equilibration time: $t_{micro} \ll t \ll t_{eq}$. It is in this regime where dynamical (aging)
scaling may be encountered, often as a result of an algebraically increasing unique length that characterizes the system \cite{Brown18,Brown19}.
Finally, for very long times with $t \lesssim t_{eq}$ the typical length saturates and approaches a maximum value
as the system approaches equilibrium.

In this publication, in contrast to previous work on interacting skyrmion systems
that mainly investigated the early stages \cite{Brown18,Brown19} of the ordering process, 
we focus on the late stages where most skyrmions are in stable positions, with a few remaining
defects. Our analysis is based on event statistics where we define an event as a change
in the number of Voronoi cells with six edges. This is achieved by computing two different histograms.
For the first histogram we record the time $t$ (measured since the initial preparation of the system) at which an event
happens and increase by one the counter $\tilde{N}(t)$ that counts the number of events happening at time $t$ since preparing
the system. We build a histogram by performing many (between 1,000 and 10,000) simulations. The quantity shown in the
figures is then the ensemble averaged number of events per skyrmion at time $t$: $N(t) = \tilde{N}(t)/(N_s n_{runs})$
where $n_{runs}$ is the total number of independent runs. For the second histogram we measure the time $\Delta t$ separating
two consecutive events and increase by one the counter $\tilde{M}(\Delta t)$ that counts the number of events separated by
the time interval $\Delta t$. Again, we discuss in the following a normalized histogram that is both 
ensemble and skyrmion averaged: $M(t) = \tilde{M}(t)/(N_s n_{runs})$.

It is convenient to bin the data in order to provide smoother curves. For the figures discussed in the following, we use a bin size
of 200 timesteps for the number $N$ of events per skyrmion, whereas for the number $M$ of consecutive events separated by a fixed time interval,
the bin size is 20. We checked that other bin sizes do not change any of our conclusions.

While our emphasis has been on event statistics and normalized histograms, we have also computed at different times the pair correlation function
\begin{equation}
\label{pcf}
g(r)=\sum\limits_{i=1}^{N_s} \sum\limits_{i\ne j} \delta \left( r - r_{ij} \right)/ (2 \pi r dr \rho N_s)
\end{equation}
with the skyrmion density $\rho = 0.1$ and the thickness $dr = 0.1$ of the ring with radius $r$. As the envelop of $g(r)-1 \sim e^{-r/\varepsilon(t)}$,
we can extract from this quantity the time-dependent correlation length $\varepsilon(t)$.

\section{Results}

It is our goal to gain a qualitative and quantitative understanding of the effects the Magnus force has on the later-time ordering behavior of
a system of interacting skyrmions. Earlier studies at intermediate times have revealed \cite{Brown18,Brown19} the presence of two different relaxation regimes,
depending on the relative strengths of the Magnus force and the thermal noise. For weak (or absent) thermal noise, we are dealing with
the Magnus-force dominated regime that is characterized by a Magnus force induced acceleration of the relaxation towards the
triangular lattice. For strong thermal noise, however, the Magnus force further enhances the effect of the noise when compared to
a system that relaxes without this force being present. In order to disentagle these two effects, we consider in the following first
the case without thermal noise before discussing the more general situation where both Magnus force and thermal noise are present.

Events being defined as changes of skyrmion positions that result in some of the Voronoi cells changing their number
of edges, these events provide a reasonable proxy for the time-dependent ordering process. The discussion
of event statistics will therefore provide insights into the late stages of the ordering of interacting skyrmion systems.

\subsection{Systems without thermal noise}

\begin{figure}
 \centering \includegraphics[width=0.6\columnwidth,clip=true]{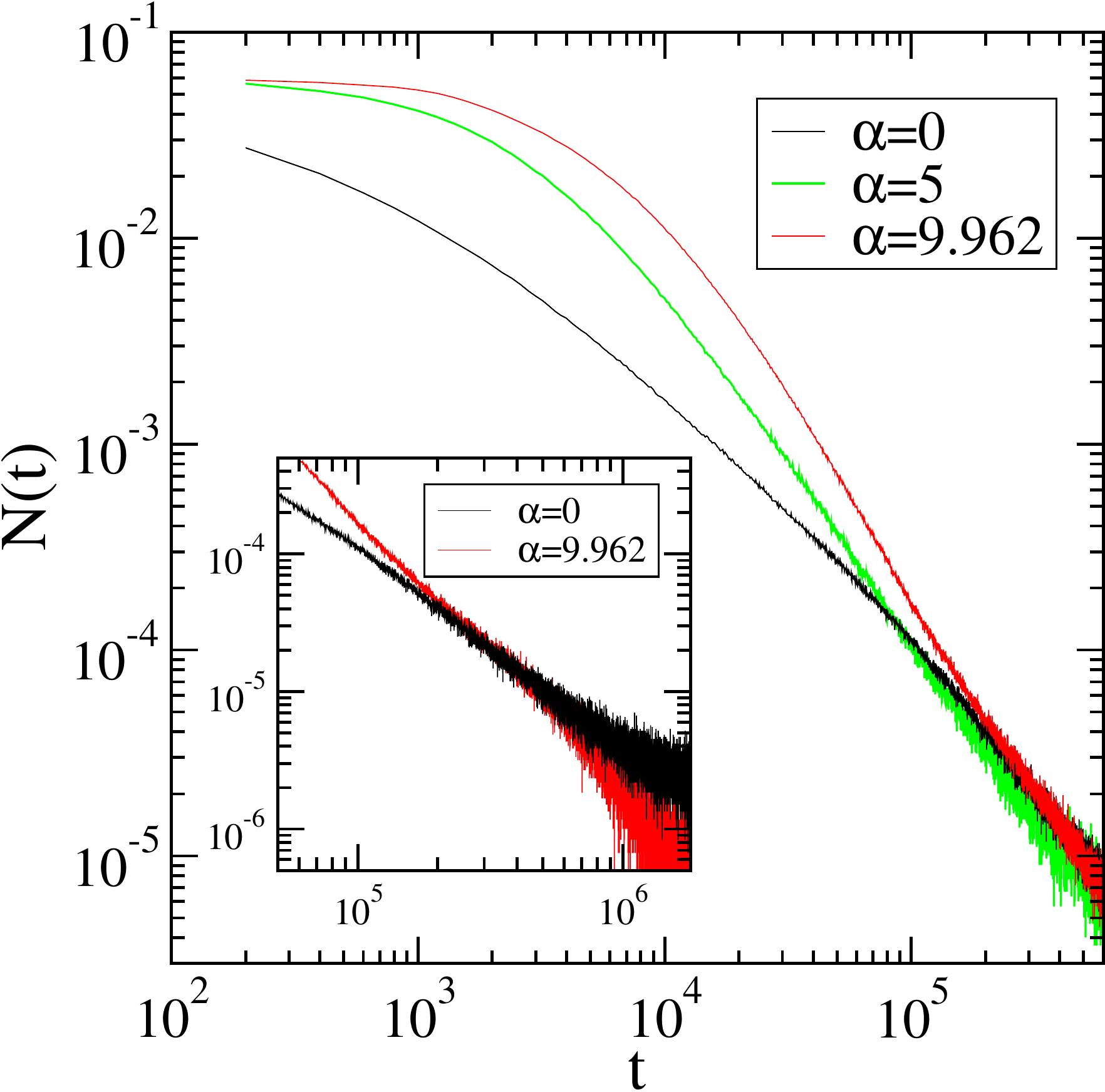}
\caption{Number $N$ of events per skyrmion as a function of the number of time steps $t$ since
preparation of the system in a disordered initial state. Data are shown for different strengths of the Magnus force $\alpha$.
The inset highlights that with Magnus force the long-time algebraic decay is governed by a different exponent than without
this force. The system parameters are $L=72$, $\lambda = 7 \xi$, and $\sigma=0$, i.e. thermal noise is absent. The data
result from an average over at least 5,000 independent runs.
}
\label{fig2}
\end{figure}

We start our discussion with Fig. \ref{fig2} that shows the ensemble averaged number of events per skyrmion as a function of the time elapsed
since preparing the system. The three curves correspond to three different values of the ratio of the Magnus force
strength and the damping coefficient studied in this work. 

%####################### Figure 2 ############################%
%####################### Figure 2 ############################%

We first note that in the short- and medium-time regimes, the number of events increases with increasing Magnus force.
This is a manifestation of the already mentioned accelerating effect the Magnus force has on the ordering process in the absence
of noise. Already, a small value of $\alpha$ accelerates relaxation, compare the green line with $\alpha = 5$ to the black 
line with $\alpha = 0$. Additional strengthening of the Magnus force mainly results in minor changes.
Independently of the value of $\alpha$, all three curves in Fig. \ref{fig2} exhibit power-law tails. This is reminiscent 
of the fat tails found in processes with rare events \cite{Taleb20} and indicates that in the late-time regime, the rare displacements
of skyrmions large enough to change the shape of Voronoi cells are algebraically distributed in time. We find that the exponent
governing this decay depends on the strength of the Magnus force, see inset in Fig. \ref{fig2}: without the Magnus force, we have 
the value $-1.42(3)$, for $\alpha=5$ the value is $-1.60(2)$, whereas for $\alpha = 9.962$ we obtain the value $-1.75(2)$. This increase of the magnitude of the exponent 
is consistent with the expectation that for larger values of $\alpha$, relaxation is faster, yielding a higher degree of order and, concomitantly,
fewer events at a fixed large time since preparation of the system.

%####################### Figure 3 ############################%
\begin{figure}
 \centering \includegraphics[width=0.6\columnwidth,clip=true]{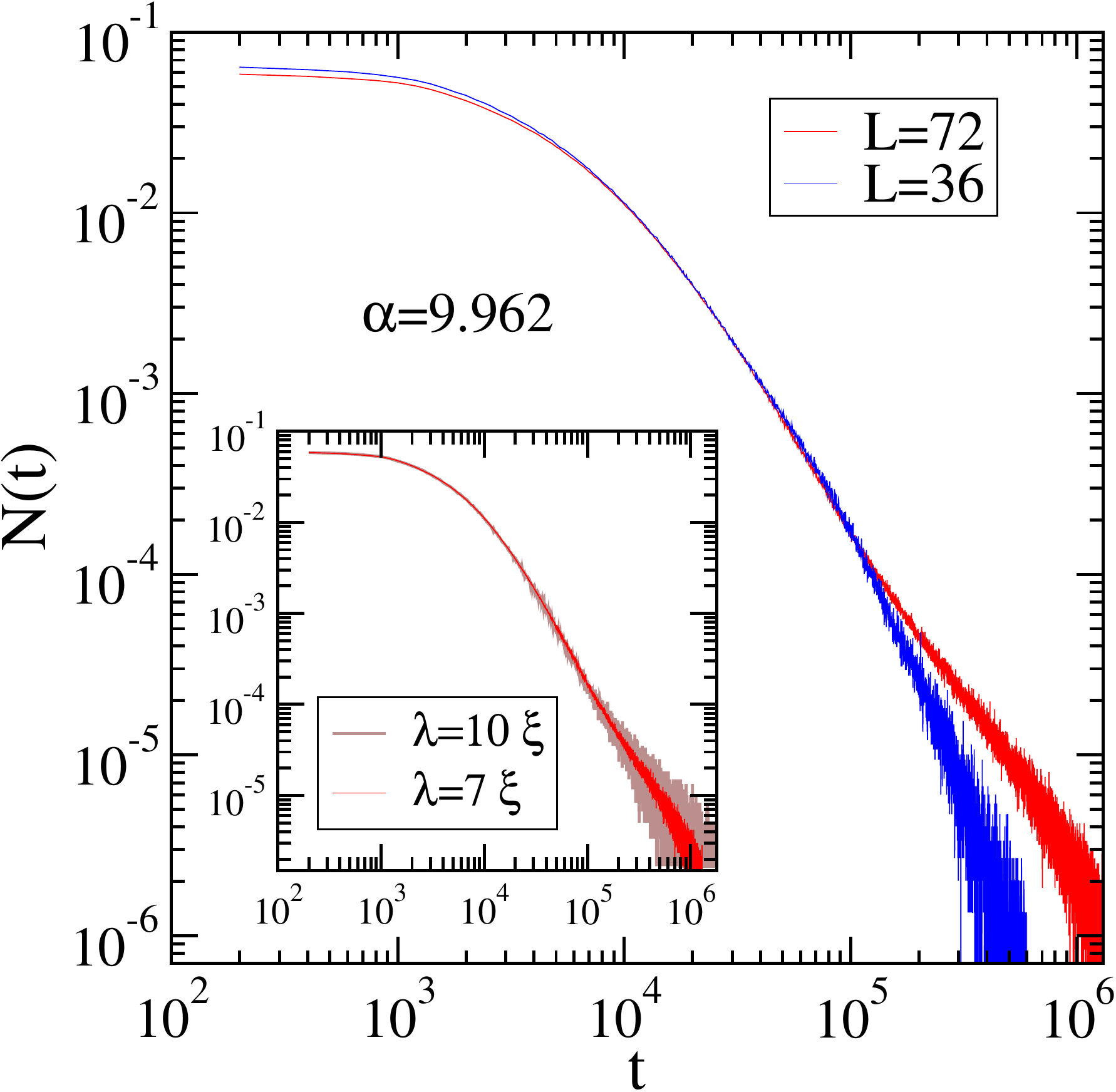}
\caption{System size dependence of the number $N$ of events per skyrmion as a function of the number of time steps $t$
for systems with the Magnus force strength $\alpha = 9.962$.
The systems have been prepared initially in a disordered state. The inset shows how $N$ changes for $L=72$ when the cut-off
length $\lambda$ is changed from $7 \xi$ to $10 \xi$. These data, which have been obtained in absence of thermal noise, result
from an average over 1,000 independent runs for $\lambda = 10 \xi$ and 10,000 independent runs for $\lambda = 7 \xi$.
}
\label{fig3}
\end{figure}
%####################### Figure 3 ############################%

Fig. \ref{fig3} illustrates the dependence of the number of events per skyrmion on two key system parameters: the system size $L$
in the main figure and the cut-off length $\lambda$ in the inset. Reducing the system size $L$ decreases the equilibration time.
The final approach to equilibrium is
revealed by deviations from a power-law behavior of $N$, and a transition to a more exponential relaxation stage (see the data
for $L=32$). As shown in the inset, increasing the cut-off length $\lambda$ (in the figure from 7 skyrmion radii to 10 skyrmion radii)
increases the noise in the ensemble averaged data and makes it harder to obtain reliable data for large times $t$, but does otherwise not affect the relaxation process. 

%####################### Figure 4 ############################%
\begin{figure}
 \centering \includegraphics[width=0.6\columnwidth,clip=true]{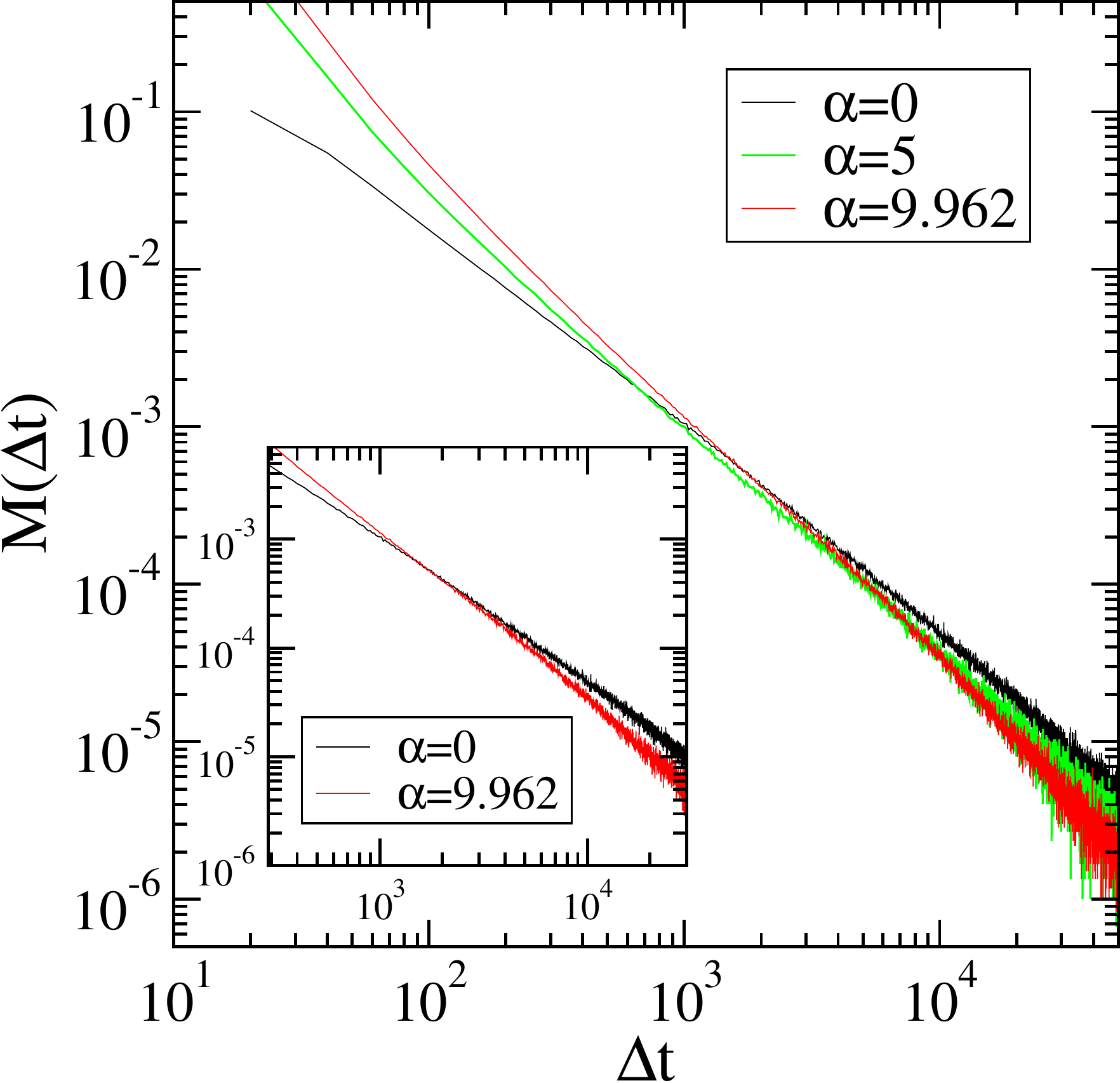}
\caption{Number $M$ of consecutive events per skyrmion separated by a fixed number of time steps $\Delta t$.
Data are shown for different strengths of the Magnus force $\alpha$.
The inset highlights that with the Magnus force, the algebraic decay of this quantity for large separating times
is governed by a different exponent than without
this force. The system parameters are $L=72$, $\lambda = 7 \xi$, and $\sigma=0$, i.e. thermal noise is absent. The data
result from an average over at least 5,000 independent runs. All the systems have been prepared initially in a disordered state.
}
\label{fig4}
\end{figure}
%####################### Figure 4 ############################%

We also analyzed how the number $M$ of consecutive events changes as a function of the time $\Delta t$ elapsed between these two consecutive events, see Fig. \ref{fig4}.
For large separation times between consecutive events, $M$ decays algebraically, and this is both in the presence and the absence of a Magnus force. The strength of the Magnus force 
impacts the exponent of the algebraic decay in a similar way as for the quantity $N$: the system orders quicker for a larger Magnus force, and the probability of encountering
two consecutive events separated by a large time difference is reduced. This is captured in the values of the power-law exponent: $-1.56(1)$ without Magnus force, whereas
with Magnus force we have $-1.75(2)$ for $\alpha = 5$ and $-1.84(3)$ for $\alpha = 9.962$. Additional differences between the cases with and without Magnus force also show
up for small separation times between consecutive events, as the quicker relaxation in presence of the Magnus force results in a larger number of events that happen close
together.

%####################### Figure 5 ############################%
\begin{figure}
 \centering \includegraphics[width=0.6\columnwidth,clip=true]{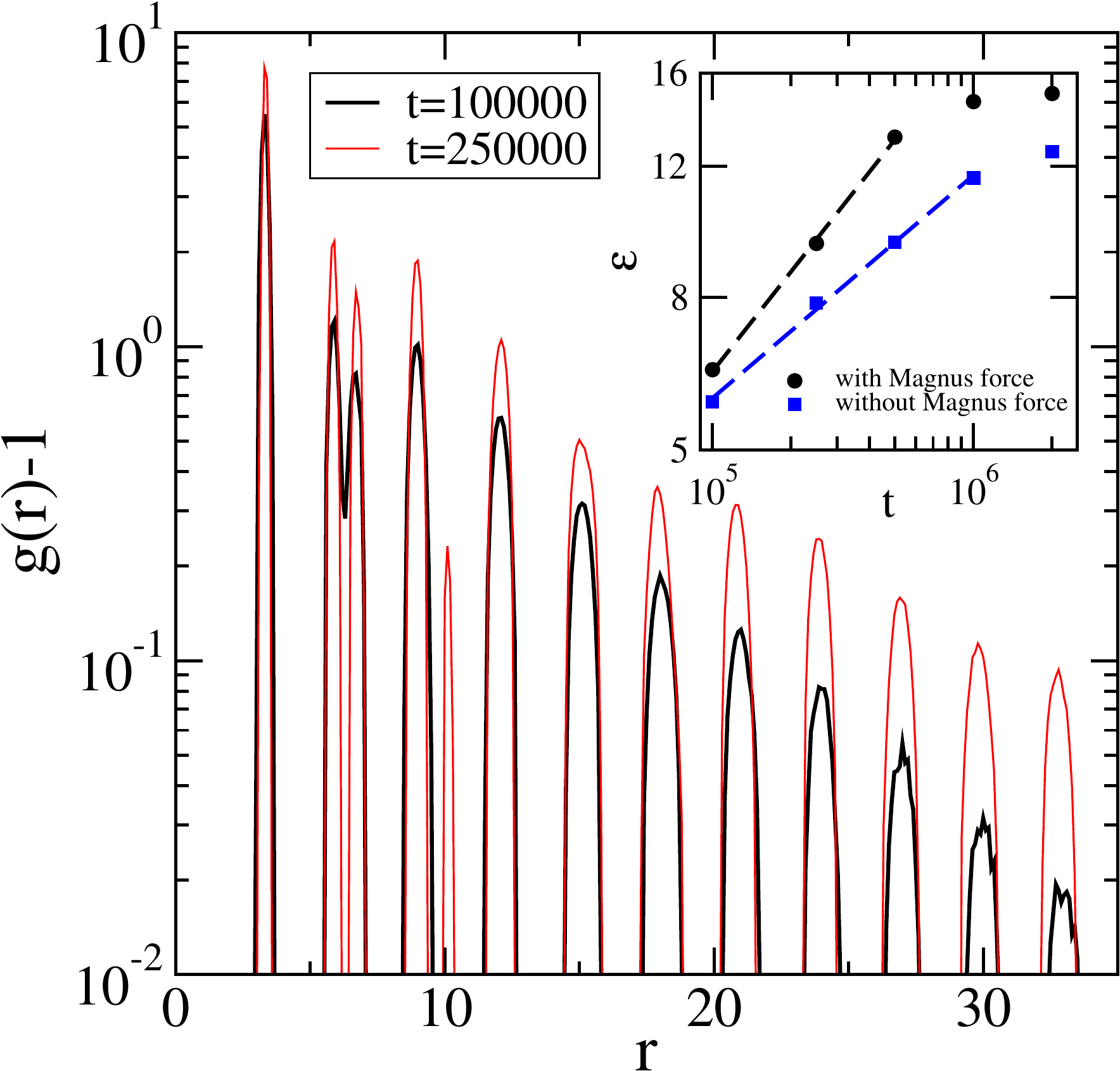}
\caption{Pair correlation function $g(r)$ as a function of distance $r$ for two different times with the Magnus force but
without noise. Plotting $g(r)-1$ in a linear-log plot shows that the envelop
decays exponentially with a time-dependent correlation length $\varepsilon(t)$. The inset compares the correlation lengths 
with and without Magnus force in absence of noise. The dashed lines are power-law fits with exponents 0.44 for the case with Magnus force, 
whereas without Magnus force the exponent is 0.30. 
The system parameters are $L=72$, $\lambda = 7 \xi$, and $\sigma=0$, i.e. thermal noise is absent. The data
result from an average over 500 independent runs. All the systems have been prepared initially in a disordered state.
}
\label{fig5}
\end{figure}
%####################### Figure 5 ############################%

Fig. \ref{fig5} shows that the pair correlation function (\ref{pcf}) evolves as a function of time, as expected for a system out of equilibrium
that undergoes an ordering process. The envelop of $g(r)-1$ displays an exponential decay as a function of distance $r$. The correlation
length extracted from the exponentially decaying envelop is shown in the inset of Fig. \ref{fig5} in presence and in absence of the Magnus force.
Focusing on the intermediate and late-time regimes, we see that the correlation length increases algebraically with time in the intermediate regime,
with an exponent 0.44 in presence of the Magnus force that is larger than the exponent 0.30 obtained in absence of the Magnus force. This again
highlights that the Magnus force accelerates the ordering process, as revealed by a larger power-law exponent, and a subsequent earlier
crossover to the late-time non-algebraic growth regime. These findings are similar to those obtained earlier \cite{Brown18} for the time-dependent
average nearest-neighbor distance, with the notable exception that the deviation of $\varepsilon(t)$ from a power law is more gradual, which allows
an easier determination of the growth exponent.

\subsection{Systems with thermal noise}

%####################### Figure 6 ############################%
\begin{figure}
 \centering \includegraphics[width=0.6\columnwidth,clip=true]{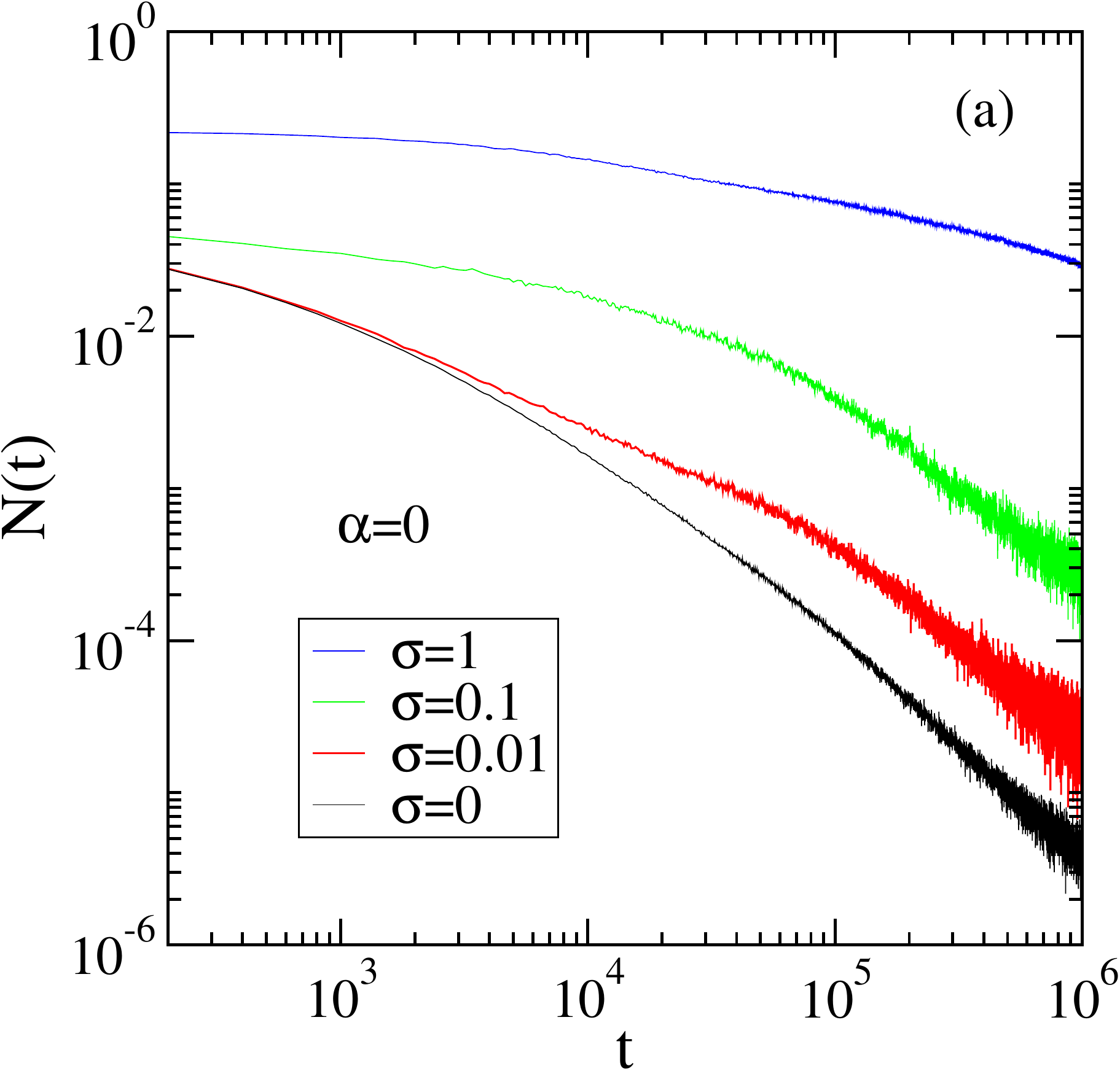}\\
 \centering \includegraphics[width=0.6\columnwidth,clip=true]{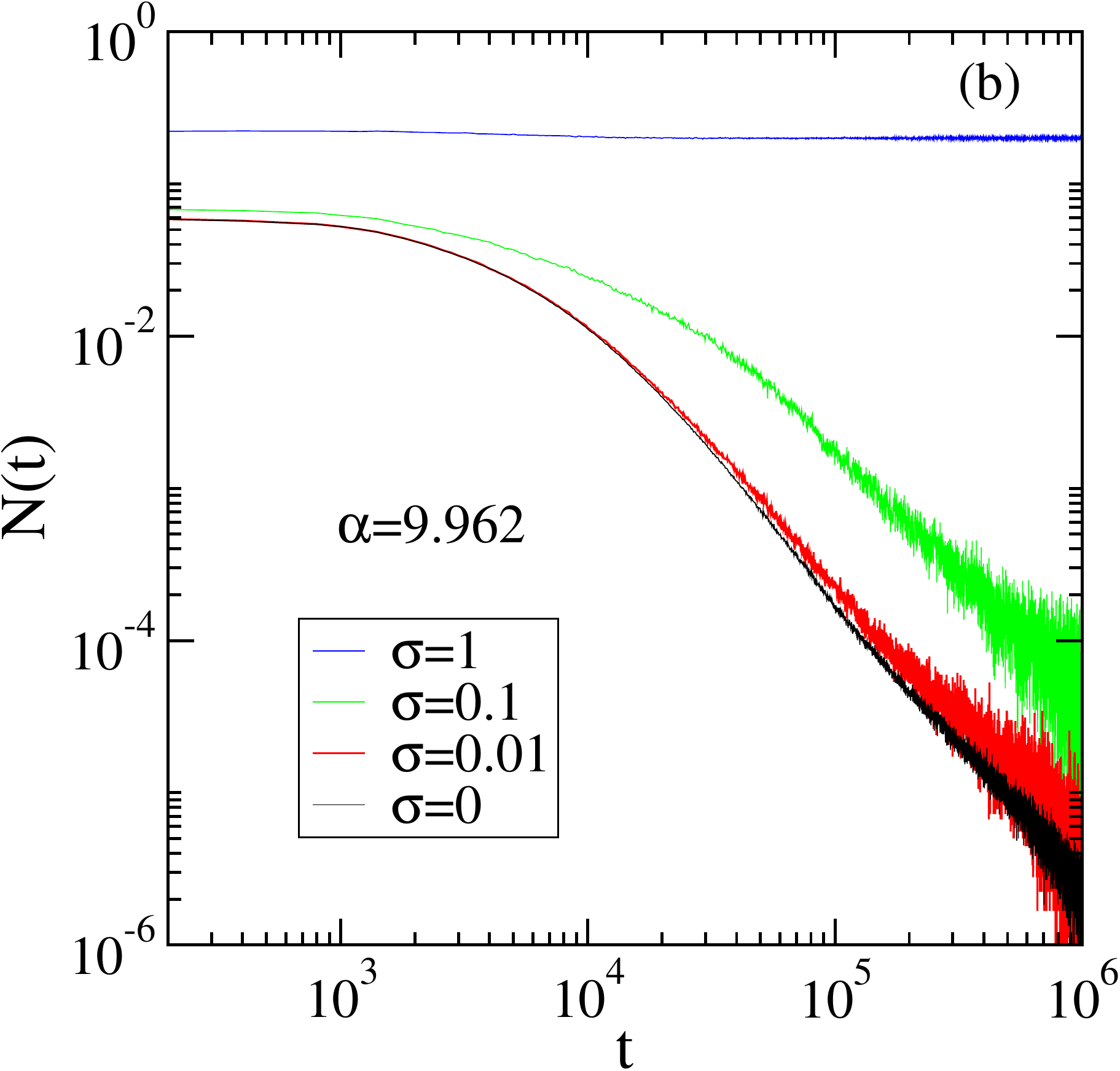}
\caption{Number $N$ of events per skyrmion as a function of the number of time steps $t$ for various
strengths of the thermal noise. (a) Systems without Magnus force, (b) systems with a Magnus force of strength
$\alpha = 9.962$. Other system parameters are set at $L=72$ and $\lambda = 7 \xi$.
The data are the average of at least 1,000 independent runs.
}
\label{fig6}
\end{figure}
%####################### Figure 6 ############################%

\begin{figure}
 \centering \includegraphics[width=0.6\columnwidth,clip=true]{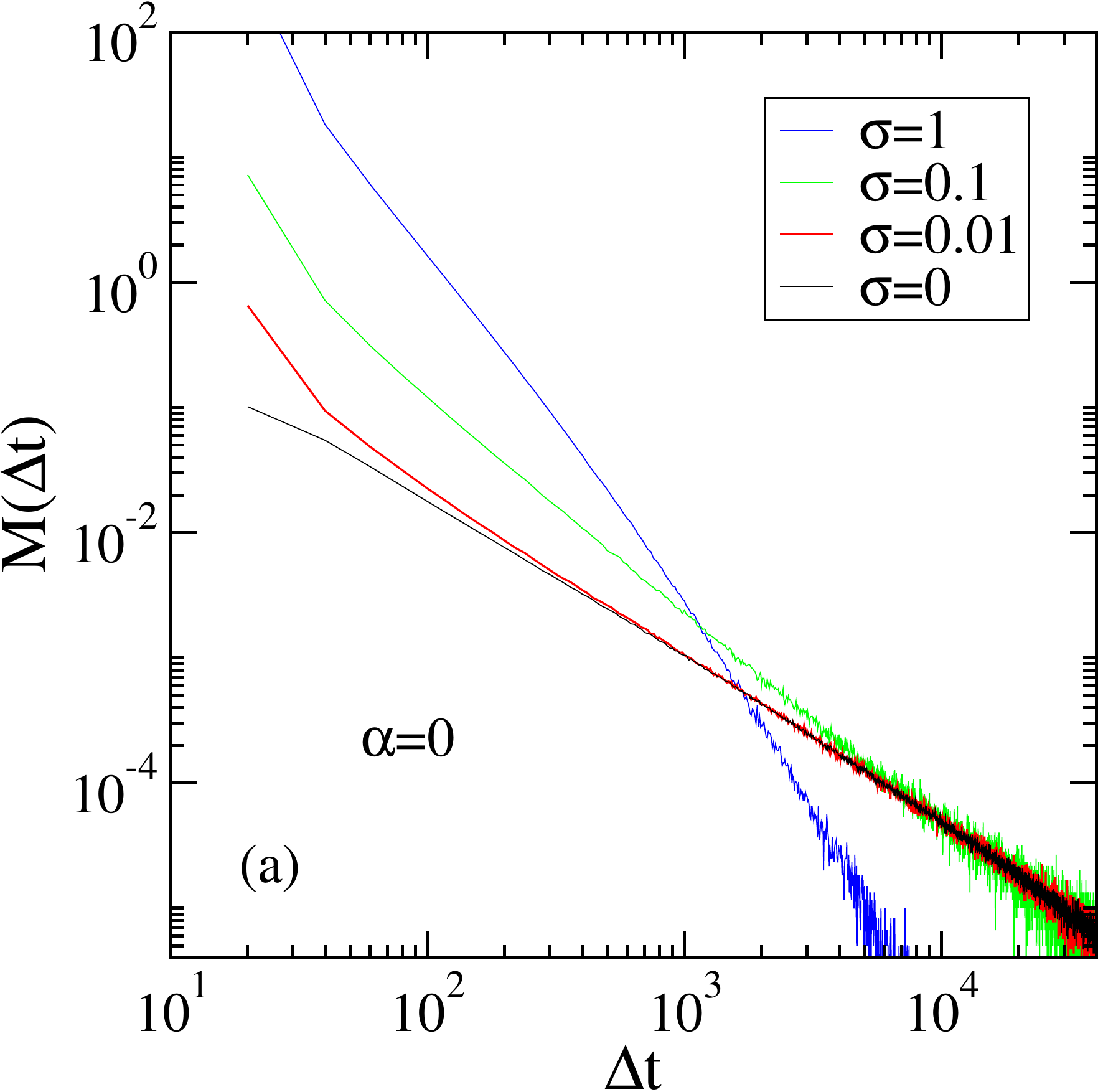}\\
 \centering \includegraphics[width=0.6\columnwidth,clip=true]{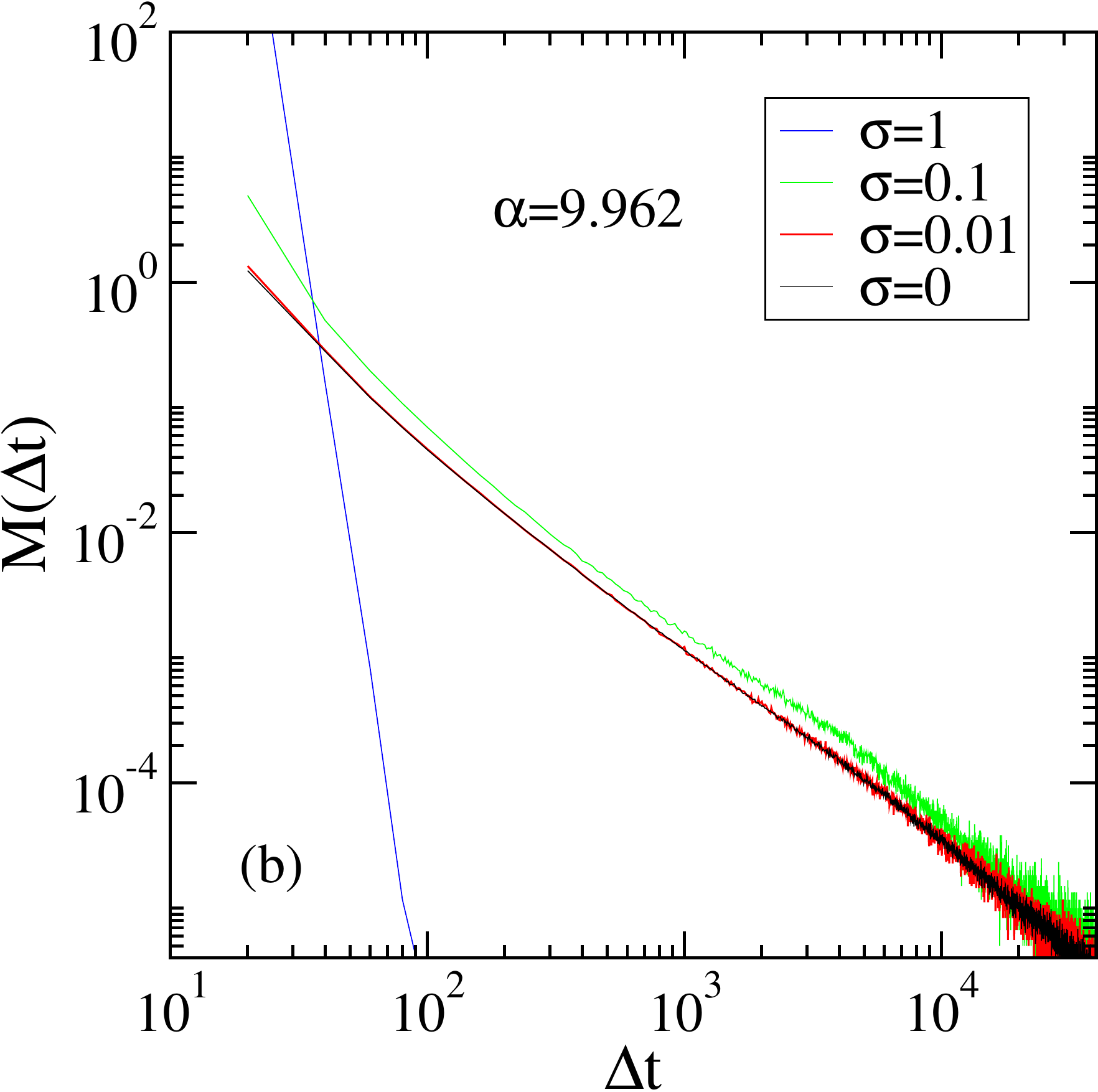}
\caption{Number $M$ of consecutive events per skyrmion separated by a fixed number of time steps $\Delta t$
for various strengths of the thermal noise. (a) Systems without Magnus force, (b) systems with a Magnus force of strength
$\alpha = 9.962$. Other system parameters are set at $L=72$ and $\lambda = 7 \xi$.
The data are the average of at least 1,000 independent runs.
}
\label{fig7}
\end{figure}

The effect of noise on the relaxation process is shown in Figures \ref{fig6} and \ref{fig7}. Inspection of these two figures reveals a variety of aspects that
need to be mentioned. In general, adding noise results in increasing the total number of events at a given time since preparation of the system. Particles are more
mobile due to the random kicks received because of thermal noise which results in an overall increase of skyrmion displacements that change the number of edges of some cells.
In absence of the Magnus force, see Fig. \ref{fig6}a, already a small value of the noise strength
yields pronounced deviations from the $\sigma =0$ data. For $\alpha = 9.962$, however, the same noise level results in only minor modifications, as shown in 
Fig. \ref{fig6}b. A strong Magnus force and the resulting curved trajectories of the skyrmions diminish the effects of weak random kicks. For large noise strengths, as for example
$\sigma =1$ in the figure (blue lines), the reverse situation is encountered, with the noise having a more pronounced effect in presence of the Magnus force. Indeed, for $\sigma =1$
we are in the noise-dominated regime \cite{Brown18,Brown19} where the Magnus force enhances the effects coming from thermal noise. Consequently, in Fig. \ref{fig6}b,
the number of events per skyrmions is, to a large extent, independent of the time since preparing the system, see blue line. This is different in Fig. \ref{fig6}a
for $\alpha =0$ and $\sigma =1$, where a slow ordering process persists that results in a slow decrease of the blue line.

We note that for small noise strengths $\sigma \le 0.1$ the algebraic decay at large values of $t$ is largely independent of the noise strength, yielding 
values for the exponent that are consistent with those obtained for $\sigma = 0$, both in the absence and presence of the Magnus force. The much enhanced noise in the
data makes a precise measurement of the exponent in presence of thermal noise difficult.

%####################### Figure 7 ############################%
%####################### Figure 7 ############################%

The curves shown in Fig. \ref{fig7} for the number of consecutive events per skyrmion for a fixed time difference provide a complementary view of the effects of thermal noise.
This quantity also reveals the reduced impact that weak thermal noise has on the ordering process in the presence of the Magnus force, including the fact that the exponents of
the power-law decay are not changed when adding weak noise. For strong thermal noise, two consecutive events
separated by more than a hundred timesteps are not encountered in the presence of the Magnus force, see Fig. \ref{fig7}b, which again highlights that the Magnus force reinforces 
the disordering effect of strong thermal noise.

\section{Conclusion}
Up to now the systematic numerical investigation of the dynamic relaxation processes in interacting skyrmion systems have been restricted to the aging scaling
regime encountered at intermediate times \cite{Brown18,Brown19}, i.e. at times that are large compared to microscopic time scales but small compared to the equilibration time.
This work presents, to our knowledge for the first time, an analysis of the properties of skyrmion matter at late stages of the ordering process, where 
skyrmion displacements that enhance locally ordering happen rarely. In order to gain insights into this late-stage relaxation process, we propose to
apply event statistics to skyrmion systems.

The Magnus force, a velocity dependent force that acts perpendicular to the direction of propagation and therefore yields curved trajectories, has a major impact on
ordering processes taking place in systems of interacting skyrmions \cite{Brown18,Brown19}. In the absence of external drive, one identifies two different dynamic
regimes, depending on the relative strengths of the Magnus force and the thermal noise. For weak thermal noise, the system is in the Magnus-force dominated regime where
the Magnus force yields an acceleration of the ordering process when compared to the case without the Magnus force. The noise-dominated regime prevails
for strong thermal noise and is characterized by the fact that the Magnus force enhances the disordering effects of the noise.

In this paper, we have focused on the ordering process of skyrmion matter without external drive and in situations where interactions with pinning defects can be 
neglected. Through the study of the number of events (defined as rearrangements that change the edge numbers of some cells obtained through Voronoi tesselation)
at a fixed time since preparing the systems, as well as of the number of consecutive events happening for a fixed time interval, interesting information on the relaxation process
are obtained. The two dynamic regimes yield a characteristic behavior of these quantities that can easily be identified. In the noise-dominated regime, both
quantities display power-law decays governed by exponents whose values depend on the strength of the Magnus force. Interestingly, these exponents are found
to be unchanged in the presence of weak noise, so that the counting of events as defined in this work may provide a possible way to investigate the different
dynamic regimes in experimental settings as those discussed recently in \cite{Zaz20}.

In cases where external drive and / or pinning defects are present, more complicated dynamic scenarios are encountered away from stationarity.
For example, strong attractive pins can capture a substantial fraction of the skyrmions, which then results in caging effects for the remaining skyrmions due to 
the repulsive skyrmion-skyrmion interactions \cite{Brown19}. We expect that in this and related scenarios the number of events per skyrmion obtained from a Voronoi
tesselation displays complicated, but characteristic, features that are worth studying in the future.

\begin{acknowledgments}
This material is based upon work supported by the U.S. Department of Energy, Office of Science, Office of Basic Energy Sciences, 
Division of Materials Sciences and Engineering under Award Number DE-SC0002308.
We thank Uwe C. T\"{a}uber for a useful discussion.
\end{acknowledgments}

\end{document}